\documentstyle[prb,aps,twocolumn,psfig,floats]{revtex}
\begin{document}
\draft
\title{Two-impurity Kondo problem for correlated electrons}
\author{Karen Hallberg$^1$ and Reinhold Egger$^2$}
\address{${}^1$Max-Planck-Institut f\"ur Physik komplexer Systeme,
Bayreuther Str.~40, D-01187 Dresden, Germany\\
${}^2$Fakult\"at f\"ur Physik, Albert-Ludwigs-Universit\"at,
Hermann-Herder-Stra{\ss}e 3, D-79104 Freiburg, Germany}
\date{Date: \today}
\maketitle
\begin{abstract}
The behavior of two magnetic impurities coupled to correlated
electrons in one dimension is studied using the DMRG technique
for several fillings.
On-site Coulomb interactions among the electrons
lead to a small Kondo screening cloud
and an overall suppression of magnetic order. 
For arbitrary electronic correlations and 
large inter-impurity distances $R$,
we find a $1/R^2$ decay of magnetic correlations.
\end{abstract}

\narrowtext
The almost complete understanding of the behavior of
magnetic impurities interacting with conduction electrons 
in metals is one of the major achievements of modern condensed
matter theory.  The impurity magnetism is mainly affected
by two distinct mechanisms, namely Kondo screening\cite{hewson} and the 
Ruderman-Kittel-Kasuya-Yosida (RKKY) interaction.\cite{rkky}
While the RKKY interaction typically favors magnetic ordering
of the impurities, the Kondo effect works against order 
since it tends to quench individual impurity spins. This competition
characterizes the magnetism in many heavy-fermion compounds,\cite{hewson}
where one essentially deals with a Kondo lattice of  rare-earth impurities
carrying a local moment due to inner $f$-electrons.
As emphasized by Varma,\cite{varma}
the basic features of the interplay between the Kondo
effect and the RKKY interaction are already contained in
the two-impurity Kondo problem which has attracted
much attention lately.\cite{two}

So far all studies of the two-impurity Kondo problem
have ignored correlations among the conduction
electrons.  While this is certainly
a reasonable assumption in most conventional metals, it breaks down
if deviations from Fermi liquid theory become important.
For instance,  Brugger {\em et al.}\cite{brugger} reported experimental
evidence for  heavy-fermion behavior in the strongly correlated material
Nd$_{2-x}$Ce$_x$CuO$_4$.
This behavior has been modeled by considering a lattice of 4f 
Nd ions interacting with strongly correlated conduction
electrons in the copper-oxide planes.\cite{fulde}
Furthermore, experimental studies of
magnetic impurities in one-dimensional (1D) quantum wires
(which are described by the strongly
correlated  Luttinger liquid state\cite{schulz})
are coming into reach due to recent fabrication advances.\cite{yacoby}

In this paper, as a prototypical example for the effects of 
correlations among conduction electrons, we employ
the density matrix renormalization group (DMRG) method \cite{dmrg}
to study the case of 1D interacting electrons. 
Recent DMRG studies for the Kondo lattice model
have investigated the influence of electronic correlations on
charge or spin gaps.\cite{lattice}
Here we present results for  the simpler and physically
more transparent two-impurity Kondo model, with main focus
on the impurity spin-spin correlations.

In the absence of spin or lattice instabilities, 1D interacting electrons
form a Luttinger liquid characterized by a dimensionless interaction
strength $g$. Generally, for repulsive interactions, one has $0<g<1$,
and $g=1$ is the noninteracting (Fermi liquid) value. While the
Luttinger liquid has been thoroughly studied for the 
clean case and in the presence of
elastic potential scattering,\cite{schulz}  the implications
of magnetic impurities are only beginning to emerge.
On the one hand, the Kondo effect 
in a Luttinger liquid\cite{kondo}  leads to the formation 
of a ground-state many-body singlet. The impurity spin is
screened by the electrons like in the conventional single-channel
Kondo effect for noninteracting electrons, albeit with a 
larger Kondo temperature. One finds $T_K
\sim (\rho_0 J)^{2/(1-g)}$ instead of the uncorrelated
result $T_K\sim \exp(-1/\rho_0 J)$, where $J>0$ is the 
antiferromagnetic exchange coupling and $\rho_0$ the
electronic density of states at the Fermi energy.
On the other hand, the  RKKY interaction 
becomes quasi-long-ranged in a Luttinger liquid.
For two spin-$\frac12$ impurities described
by spin operators $\bf{S}_1$ and ${\bf S}_2$, one finds the effective
 RKKY Hamiltonian,\cite{egger}
\begin{equation}\label{rk}
H_{\rm eff} = {\cal K}\, {\bf S}_1  {\bf S}_2 \;, \qquad {\rm where}\;
\; {\cal K} \sim J^2 R^{-g} \cos(2 k_F R)\;.
\end{equation}
The $2k_F$-oscillatory RKKY indirect exchange coupling
${\cal K}$ is defined by second-order perturbation
theory in the exchange coupling $J$. Higher-order terms
(which are also responsible for the Kondo effect)
cannot be written in the simple form (\ref{rk}).
 As a function of inter-impurity distance $R$, the RKKY coupling 
decays slower than the conventional $1/R$ Fermi liquid result,
which is recovered  by putting $g=1$.

\begin{figure}[t]
\unitlength0.75cm
\begin{picture}(11,3)
\thinlines
\put(0,2.5){\line(1,0){11}}
\put(4,1.25){\line(0,1){1.25}}
\put(7,1.25){\line(0,1){1.25}}
\thicklines
\put(0,2.5){\circle*{0.3}}
\put(1,2.5){\circle*{0.3}}
\put(2,2.5){\circle*{0.3}}
\put(3,2.5){\circle*{0.3}}
\put(4,2.5){\circle*{0.3}}
\put(5,2.5){\circle*{0.3}}
\put(6,2.5){\circle*{0.3}}
\put(7,2.5){\circle*{0.3}}
\put(8,2.5){\circle*{0.3}}
\put(9,2.5){\circle*{0.3}}
\put(10,2.5){\circle*{0.3}}
\put(11,2.5){\circle*{0.3}}
\put(4,1){\circle{0.5}}
\put(7,1){\circle{0.5}}
\put(3.5,1.6){$J$}
\put(7.3,1.6){$J$}
\end{picture}
\caption[]{\label{fig1} 
Two spin-$\frac12$ impurities (large open circles)
coupled to a 1D Hubbard chain (filled circles).
For this example, the distance of the impurities is $R=3$,
and $N=12$.}
\end{figure}
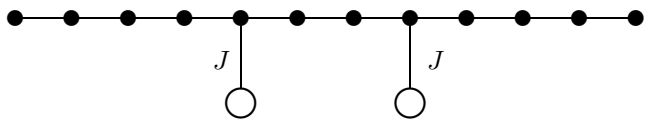

The competition between the RKKY interaction and the Kondo
effect shows up in the magnetic correlations between the
impurities, $\langle {\bf S}_1  {\bf S}_2 \rangle$.  From Eq.~(\ref{rk}),
one would find\cite{fye}
\begin{equation}\label{ss}
\langle {\bf S}_1  {\bf S}_2 \rangle = \frac{3}{4} \,
\frac{1- \exp({\cal K}/k_B T)}{3+\exp({\cal K}/k_B T)} \;.
\end{equation}
Due to higher-order contributions in the exchange coupling $J$,
this expression is expected to
saturate for temperatures below the Kondo temperature $T_K$.
Therefore the ratio between the relevant energy scales,
 ${\cal K}/T_K$, is crucial for 
 the ground-state value of $\langle {\bf S}_1 
 {\bf S}_2 \rangle$.
The sign of $\langle {\bf S}_1  {\bf S}_2 \rangle$ depends on the sign 
of ${\cal K}$ in Eq.~(\ref{rk}), and therefore 
$\langle {\bf S}_1  {\bf S}_2 \rangle$ is also $2 k_F$-oscillatory.
For very small $J$, the RKKY interaction dominates
such that  $\langle {\bf S}_1  {\bf S}_2 \rangle$ takes its maximum 
(singlet or triplet) value.
For very large $J$, Kondo screening is effective in quenching the 
impurity spins and magnetic order disappears, i.e.,
$\langle {\bf S}_1  {\bf S}_2 \rangle\to 0$. 
In order to describe the crossover from RKKY to Kondo-dominated behavior,
a non-perturbative approach is mandatory.
We note that the unstable fixed point separating these behaviors
(see Ref.\onlinecite{two}) is absent in the correlated case 
since the electron-hole symmetry required for its presence is broken. 

\begin{figure}
\centerline{
\psfig{width=8.5truecm,bbllx=60pt,bblly=55pt,bburx=580pt,bbury=700pt,angle=-90,file={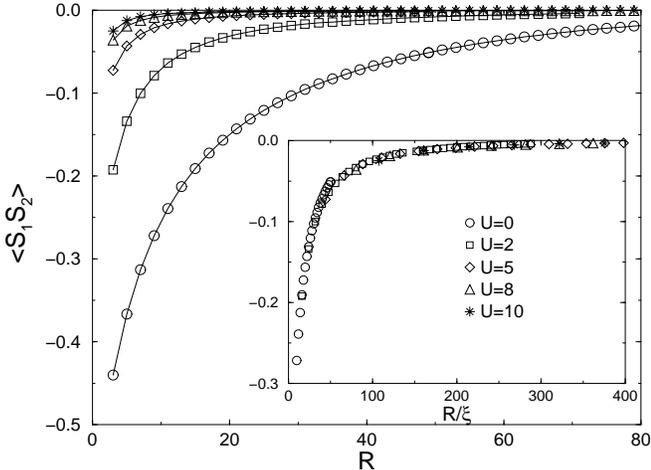}}}
\caption{Numerical results for
$\langle {\bf S}_1  {\bf S}_2 \rangle$
for $J=1$ and several $U$ at half-filling. The impurity 
distance is fixed at $R=N-1$. The inset shows scaled data,
where the distance between the impurities is
 measured in units of the lengthscale $\xi$ given in Fig.~3
The solid curves are guides to the eye only.
}
\label{fig2}
\end{figure}

We have studied two spin-$\frac12$ impurities  ${\bf S}_i$ $(i=1,2)$
coupled to the electron spin density ${\bf s}(x_i)$ 
at the respective impurity location 
by the standard exchange term 
$J \,{\bf s}(x_i)  {\bf S}_i$.  The 1D correlated
electrons are modelled as a Hubbard chain, with nearest-neighbor hopping
$t=1$ and repulsive on-site Coulomb interaction $U\geq 0$. 
The low-temperature behavior of the Hubbard chain away 
from half-filling is equivalent to a Luttinger liquid, and
the appropriate values for $g$ as a function of $U$ and the filling factor
can be found in Ref.\onlinecite{schulz90}.
Directly at half-filling, a finite $U$ causes a charge gap leading
to a Mott insulator instead of the metallic Luttinger liquid
state.  However, since the spin sector remains
gapless, the magnetic properties of the Hubbard chain
still follow the Luttinger liquid predictions, but with
$g=0$. Away from half-filling, one has
$1/2 \leq g \leq 1$. We have primarily 
studied the half-filled case, for which
 DMRG is known to be very accurate.\cite{dmrg} 
To demonstrate that similar conclusions apply to
the metallic ($g>0$) Luttinger liquid, we have 
also investigated the quarter-filled case,
for which some data are shown below. 

For accuracy reasons, only DMRG results using the finite-lattice 
version\cite{dmrg} and open boundary conditions are presented.
As the appropriate measure of magnetic order, the ground-state value of 
$\langle {\bf S}_1  {\bf S}_2 \rangle$ has been calculated.
We included $m\simeq 200$ states of the reduced density
matrix to study chains of even length  $N\simeq 80$, 
with an accuracy better than 1\% for $\langle {\bf S}_1  {\bf S}_2 \rangle$ 
even in unfavorable cases (small $U$ and quarter-filling).
The two impurities are arranged symmetrically around the center
of the chain. In units of the lattice spacing, their
distance $R$ takes odd values varying  
 between 1 (impurities at the chain center)
and $N-1$ (impurities at opposite chain ends).
This geometry is shown in Fig.~\ref{fig1}.
The bulk behavior is then reproduced in the limit
of large $N$ and $R \ll N/2$. 

\begin{figure}
\centerline{
\psfig{width=8.5truecm,bbllx=60pt,bblly=55pt,bburx=580pt,bbury=700pt,angle=-90,file={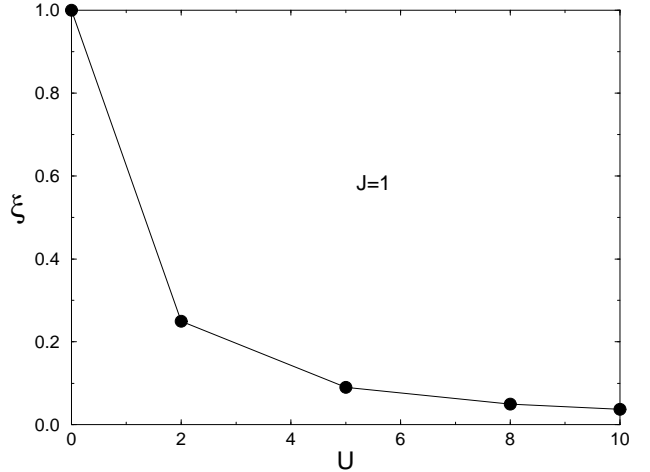}}}
\caption{ Lengthscale $\xi$ used 
in Fig.~2 as a function of $U$. 
We have chosen $\xi=1$
for $U=0$, and then  determined $\xi$ by matching data onto the 
scaling curve. The solid curve is a guide to the eye only.
}
\label{fig3}
\end{figure}

In Fig.~\ref{fig2}, DMRG  results for 
 $\langle {\bf S}_1  {\bf S}_2 \rangle$
are shown for $J=1$ and various $U$.  Since the data were obtained
 for half-filling and  $R$ is odd, RKKY correlations 
are always antiferromagnetic such that
 $\langle {\bf S}_1  {\bf S}_2 \rangle<0$.   We notice several
features in Fig.~\ref{fig2}: (a) Increasing the Coulomb interaction $U$
leads to smaller values of $|\langle {\bf S}_1  {\bf S}_2 \rangle|$.
(b) The magnetic correlations $\langle {\bf S}_1  {\bf S}_2 \rangle$
fall off with increasing distance due to 
the decreasing RKKY coupling ${\cal K}$. (c)  As shown in the
inset, all data points can be scaled onto a universal curve 
by using the lengthscale $\xi$ shown in Fig.~\ref{fig3}.
The overall behavior of $\xi$ is quite similar to the Kondo screening
length $v_F/T_K$, which, roughly speaking, determines the extent of
the electronic screening cloud around the 
impurity.\cite{hewson,egger,sorensen}
Since $T_K$ increases with correlations, the observed decrease
of $\xi$ with increasing $U$ is in accordance with the results of 
Ref.\onlinecite{kondo}. In that sense, 
electronic correlations imply a smaller screening cloud.

\begin{figure}
\centerline{
\psfig{width=8.5truecm,bbllx=60pt,bblly=55pt,bburx=580pt,bbury=700pt,angle=-90,file={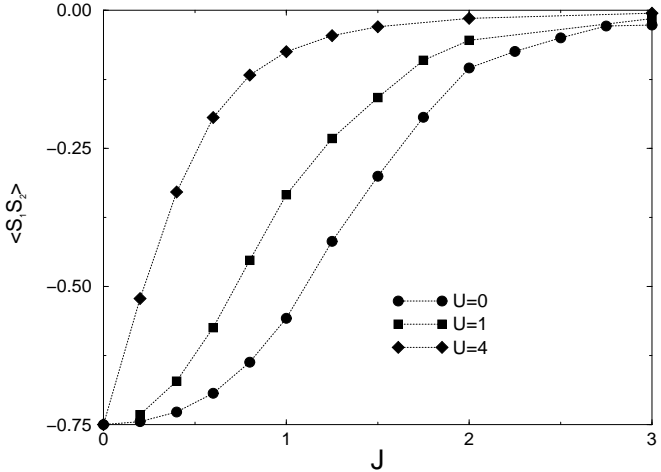}}}
\caption{$\langle {\bf S}_1  {\bf S}_2 \rangle$ as a function of  $J$ 
at half-filling for $R=5$  and various $U$. 
 Dotted curves are guides to the eye only.
}
\label{fig4}
\end{figure}

As depicted in Fig.~\ref{fig4}, DMRG results  for the
$J$-dependence of $\langle {\bf S}_1  {\bf S}_2 \rangle $
at the fixed distance $R=5$ explicitly
demonstrate the competition between RKKY and Kondo-dominated
behavior. Since  almost the same curves were found for $N=30$,
finite-size effects seem to play only a very minor role here.
For small $J$, the (here antiferromagnetic)
RKKY interaction leads to a singlet, 
$\langle {\bf S}_1  {\bf S}_2 \rangle \to -3/4$, while the Kondo effect
destroys magnetic order for large $J$ such that 
$\langle {\bf S}_1  {\bf S}_2 \rangle\to 0$. 
The qualitative influence of Coulomb correlations can be
read off from Figs.~\ref{fig2} and \ref{fig4}.  The magnetic correlations 
$|\langle {\bf S}_1  {\bf S}_2 \rangle|$ are significantly 
reduced by switching on and increasing the Coulomb interaction $U$. Regarding
the impurity spins, {\em RKKY-related magnetic order is suppressed by 
electronic correlations}.
The same qualitative finding was obtained for quarter-filling and
other values of $U$ or $R$ under consideration. 

This behavior can be simply rationalized as follows.  Switching on the
Hubbard-$U$ implies an increase in the Kondo temperature and 
enhances the importance of Kondo screening.
Since the correlated RKKY law (\ref{rk}) does not 
change appreciably with increasing $U$
for the values of $R$ studied in
the DMRG simulations, the increase in $T_K$ is decisive here and
leads to the observed partial destruction of magnetic order.

\begin{figure}
\centerline{
\psfig{width=8.5truecm,bbllx=60pt,bblly=55pt,bburx=580pt,bbury=700pt,angle=-90,file={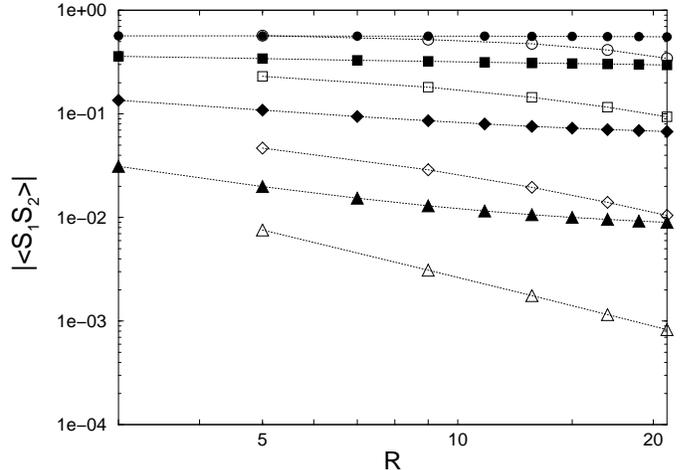}}}
\caption{Distance dependence of
 $\langle {\bf S}_1  {\bf S}_2 \rangle$ on a double-logarithmic scale
for $U=4$  and $J=0.25$ (circles), $J=0.5$ (squares), $J=1$ (diamonds), and
$J=2$ (triangles). Open symbols are for quarter-filling,
filled symbols for half-filling.
}
\label{fig5}
\end{figure}

\begin{figure}
\centerline{
\psfig{width=8.5truecm,bbllx=60pt,bblly=55pt,bburx=580pt,bbury=700pt,angle=-90,file={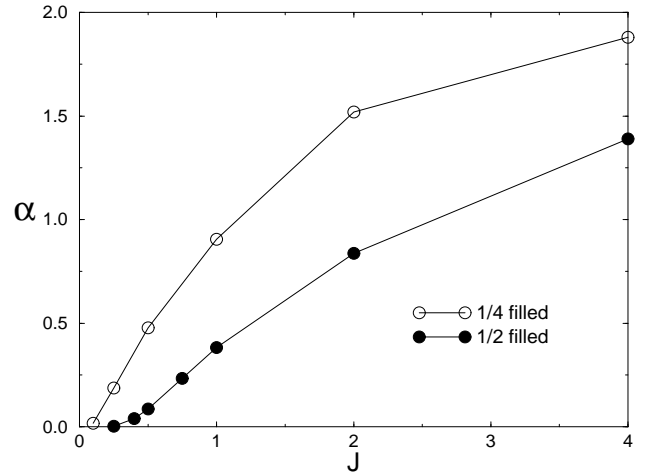}}}
\caption{ Power-law exponent $\alpha$ 
for the data in Fig.~5.
}
\label{fig6}
\end{figure}

Next we study the distance dependence of 
$\langle {\bf S}_1  {\bf S}_2 \rangle$ in the bulk limit $R\ll N/2$,
for which DMRG results are shown in Fig.~\ref{fig5}.
For the quarter-filled case and the configurations considered here (odd $R$),
the impurity correlations $\langle {\bf S}_1
{\bf S}_2 \rangle$ exhibit an oscillatory behavior.
We plot here only the negative values; similar conclusions are reached
for the positive ones.
For small enough $J$, the RKKY interaction locks the impurity 
spins in a singlet state irrespective of the distance $R$.  
Increasing $J$ causes two effects: (a) The Kondo effect leads to
smaller values of $|\langle {\bf S}_1  {\bf S}_2 \rangle|$.
(b) There is an (approximate) power-law decrease in the distance dependence
\begin{equation}
 \langle {\bf S}_1  {\bf S}_2 \rangle \sim \cos(2k_FR)/ R^\alpha\;.
\end{equation}
The exponent $\alpha$ is shown in Fig.~\ref{fig6} for the data
of Fig.~\ref{fig5}.
Clearly, $\alpha=0$ for very small $J$,
and then $\alpha$ continuously increases with $J$. For large $J$,
the limiting value $\alpha=2$ is approached for 
{\em arbitrary} $U$ and filling factor.

Since the RKKY coupling ${\cal K}\to 0$ as $R\to \infty$, 
the large-distance behavior should
effectively be determined by the large-$J$ behavior.
Therefore, from our DMRG results we expect the asymptotics
\begin{equation} \label{as}
 \langle {\bf S}_1   {\bf S}_2 \rangle \sim  \cos(2k_F R)/ R^2
\end{equation}
for any value of $J$. For $J\to \infty$, Eq.~(\ref{as}) holds on
all lengthscales. Furthermore, additional DMRG results not shown here 
reveal that the asymptotic law (\ref{as}) 
is approached faster in the case of strong electronic
correlations. 

The large-$J$ value $\alpha=2$ can be understood analytically.
For $J=\infty$, each impurity spin forms a strongly coupled
singlet with the conduction electron spin at that site (see Fig.~\ref{fig1}).
This singlet cannot be broken up, and therefore the Hubbard chain
is effectively cut at the impurity sites. 
The leading $1/J$ contribution to 
$\langle {\bf S}_1  {\bf S}_2 \rangle$ can then be computed
by  open boundary bosonization.\cite{fabrizio} To proceed, we  write 
\begin{equation} \label{11}
\langle {\bf S}_1   {\bf S}_2 \rangle  = (2\pi/k_F)^2 \;
\langle {\bf s}(x_1)   {\bf s}(x_2) \rangle  \;, 
\end{equation}
since ${\bf s}(x_i)$ is antiparallel to ${\bf S}_i$ in $1/J$ accuracy.
We then compute $\langle {\bf s}(x) 
 {\bf s}(y) \rangle$   under open boundaries
at the impurity locations (we take $x_1=0$ and 
$x_2=R$),
\begin{eqnarray*}
\langle {\bf s}(x)   {\bf s}(y) \rangle &=& 
\frac{3 k_F^2}{4\pi^2}\, \cos [2k_F (x-y) - f(2x) + f(2y)]\\
&\times& \; [ 
P(2x) P(2y)]^{(1+g)/2} \left( \left[\frac{P(x-y)}{P(x+y)}\right]^{1+g}
-1 \right) \\
&& + \; (y\to -y) \;,
\end{eqnarray*}
with the functions (we consider $k_F R\gg 1$)
\begin{eqnarray*}
P(x) &=& \left \{ 1 + \left[ \frac{2k_F R}{\pi} \sin \left(\frac{
\pi x}{2 R}\right)\right]^2 \right\}^{-1/2} \\
f(x) &=& \arctan \left[ \frac{ \sin(\pi x/R) }
{\exp(\pi/k_F R) - \cos(\pi x/R)} \right] \;.
\end{eqnarray*}
Evaluating this result for $x$ near $x_1$ and $y$ near $x_2$ 
reproduces \cite{foot3} the numerically observed behavior 
(\ref{as}). Of course, the prefactor in Eq.~(\ref{as}) depends on the 
interaction strength parameter $g$.

To conclude, we have employed DMRG simulations to study
the two-impurity Kondo problem
for interacting 1D electrons. On-site Coulomb interactions were
shown to partially destroy magnetic ordering between the
impurities. The main reason for this effect is the increase
of the Kondo temperature for correlated electrons. 
For the impurity spin-spin correlations, 
we obtain a $\cos(2k_F R)/ R^2$ behavior at large distances.

We wish to thank P. Fulde,  H. Grabert, H. Schoeller, and P. Thalmeier
for useful discussions.


\begin{references}
\bibitem{foot3} Taking directly $x=x_1$ and $y=x_2$, the
spin density correlations would vanish due to the open boundary 
conditions.
\end{references}
\end{document}